\documentclass[12pt]{article} 
\usepackage{graphics}
\title{On the structure of the pion:\\ A QCD--inspired point of view}
\author{H.C. Pauli\\
        Max--Planck Institut f\"ur Kernphysik, 
        D--69029 Heidelberg}
\date{3 November 2001} 
\begin{document}
\maketitle 
\begin{abstract}
    The effective interaction between a quark and an anti-quark
    as obtained previously with by the method of iterated resolvents
    is replaced by the so called up-down-model and applied to 
    flavor off-diagonal mesons including the pion.
    The only free parameters are those of canonical quantum chromo-dynamics
    (QCD), particularly the coupling constant and the masses of the quarks.
    
    The so obtained light-cone wave function can be used to calculate 
    the pion's form factor, particularly its mean-square radius 
    can be computed analytically.
    The results allow for the exciting conclusion that the pion 
    is built by highly relativistic constituents, in strong contrast 
    to composite systems like atoms or nuclei
    with non-relativistic constituents. 
\end{abstract}
\section{Introduction}
One of the most urgent problems in contemporary physics
is to compute the structure of hadrons in terms of their
constituents, 
% like quarks and gluons, 
based on a covariant theory such as QCD.

Among the hadrons the pion is the most mysterious particle.
I have proposed an oversimplified model,
the $\uparrow\downarrow$-model, which has many drawbacks
but the virtue of being inspired by QCD and of having the same number
of parameters one expects in a full theory:
namely the 6 flavor quark masses, the strong coupling constant (7)
and one additional scale parameter (8) originating in the murky
depth of renormalization theory.

The model is QCD-inspired by virtue of the fact, that it
is based on the full light-cone Hamiltonian
as obtained from the QCD-Lagrangian in the light-cone gauge, 
with zero-modes disregarded.
In consequence, the pion is treated on the same footing as
all other pseudo-scalar and pseudo-vector mesons.

The model should be contrasted to Lattice Gauge Calculations,
see for example \cite{Schierholz00}.
It is not generally known that LGC's have considerable
uncertainty to extrapolate their results down to such
light mesons as a pion.
It is also not generally known that lattice gauge calculations
get \emph{ always strict and linear confinement}
even for QED, where we know the ionization threshold.
The `breaking of the string', or in a more physical language,
the ionization threshold is one of the hot topics
at the lattice conferences \cite{Schilling2000}.
Moreover, in order to get the size of the pion,
thus the form factor, another generation of computers 
is required, as well as physicists to run them.

The model should be contrasted also to phenomenological approaches.
They usually do not address to get the pion.
For the heavy mesons, where they are so successful \cite{Plessas01},
phenomenological model have quite many parameters,
in any case more that the above canonical ones.
A detailed comparison and systematic discussion of the
bulky literature can however be postponed, until
we are ready to solve the full Eq.(\ref{eq:1}). 

The model should be contrasted, finally, to 
Nambu\--Jona\--Lasinio\--like models which are so successful
in accounting for isospin-aspects.
I cannot quote the huge body of literature 
but mention in passing that the NJL-models are not renormalizable,
that NJL has no relation to QCD, and that NJL deals mostly with 
the very light mesons. There is no way to treat the heavy flavors,
see also \cite{Leutwyler01}.

\section{Motivation}
The light-cone approach to the bound-state problem 
in gauge theory \cite{BroPauPin98}
aims at solving  
$H_{LC}\vert\Psi\rangle = M^2\vert\Psi\rangle$.
If one disregards possible zero modes 
and works in the light-cone gauge, 
the (light-cone) Hamiltonian $H_{LC}$ 
is a well defined Fock-space operator and given in \cite{BroPauPin98}. 
Its eigenvalues are the invariant mass-squares  $M^2$
of physical particles associated with the eigenstates
$\vert\Psi\rangle$.
In general, they are superpositions of all possible
Fock states with its many-particle configurations.
For a meson, for example, holds 
\begin{center}
\(\displaystyle  
\begin{array} {rclllllll} 
      {\vert\Psi_{\rm meson}\rangle} &=& {\sum\limits_{i}\ }
      {\Psi_{q\bar q}(x_i,\vec k_{\!\perp_i},\lambda_i)}
      {\vert q\bar q\rangle} 
  &+& {\sum\limits_{i}\ }
      {\Psi_{g g}(x_i,\vec k_{\!\perp_i},\lambda_i)}
      {\vert g g\rangle }
\\&+& {\sum\limits_{i}\ }
      {\Psi_{q\bar q g}(x_i,\vec k_{\!\perp_i},\lambda_i)}
      {\vert q\bar q g\rangle }
  &+& {\sum\limits_{i}\ }
      {\Psi_{q\bar q q\bar q }(x_i,\vec k_{\!\perp_i},\lambda_i)}
      {\vert q\bar q q\bar q \rangle}
  &+& {\dots}\ .
\end{array}\)
\end{center}
If all wave functions like $\Psi_{q\bar q}$ or $\Psi_{g g}$
are available, one can analyze hadronic structure 
in terms of quarks and gluons \cite{BroPauPin98}.

For example, one can calculate the space-like form factor 
of a hadron quite straightforwardly. As illustrated in 
Fig.~\ref{fig:LLF}, it is just a
sum of overlap integrals analogous to the corresponding
non-relativistic formula \cite{BroPauPin98}:   
\begin{equation}
   F (q^2) =
   \sum_{n,\lambda_i} \sum_a e_a \int
   \overline{\prod_i} \ {dx_i\,d^2\vec k_\perp{}_i \over 16\pi^3}
   \,\psi_{n}^{(\Lambda)*}(x_i,\vec \ell_\perp{}_i,\lambda_i)
   \,\psi_{n}^{(\Lambda)}(x_i,\vec k_\perp{}_i,\lambda_i) .
\label{eq:b1}\end{equation}
Here $e_a$ is the charge of the struck quark, $\Lambda^2\gg\vec
q_\perp^{\,2}$, and
\[
  \vec \ell_\perp{}_i \equiv \cases{
  \vec k_\perp{}_i - x_i \vec q_\perp 
+ \vec q_\perp  & \hbox{for the struck quark} \cr
  \vec k_\perp{}_i - x_i \vec q_\perp        
& \hbox{for all other partons,}\cr}
\]
with $\vec q_\perp^{\,2}=Q^2=-q^2$.
All of the various form factors
of hadrons with spin can be obtained by computing the matrix element
of the plus current between states of different initial and final
hadron helicities.

\section {The method of iterated resolvents}

Because of the inherent divergencies in a gauge field theory,
the QCD-Hamiltonian in 3+1 dimensions must be regulated
from the outset. 
One of the few practical ways is vertex 
regularization \cite{BroPauPin98,Pau98},
where every Hamiltonian matrix element, particularly those of the 
vertex interaction (the Dirac interaction proper), 
is multiplied with a convergence-enforcing momentum-dependent function.
It can be viewed as a form factor \cite{BroPauPin98}.
The precise form of this function is unimportant here,
as long as it is a function of a cut-off scale ($\Lambda$).

Perhaps one can attack the problem of diagonalizing 
the (light-cone) Hamiltonian $H_{LC}$ by DLCQ,
see for example \cite{Hil00}.
But, alternatively, it might be better to reduce the many-body problem
behind a field theory to an effective one-body problem.
The derivation of the effective interaction becomes then the key issue.
By definition, an effective Hamiltonian acts only
in the lowest sector of the theory  
(here: in the Fock space of one quark and one anti-quark). 
And, again by definition, it has the same eigenvalue spectrum
as the full problem.
I have derived such an effective interaction 
by the method of iterated resolvents \cite{Pau98}, that is
by systematically expressing the higher Fock-space
wave functions as functionals of the lower ones.  
In doing so the Fock-space is not truncated
and all Lagrangian symmetries are preserved.
The projections of the eigenstates onto the 
higher Fock spaces can be retrieved
systematically from the $q\bar q$-projection, 
with explicit formulas given in \cite{Pau99b}.

%-------------------------------------------------------------------
\begin{figure} [t]
\begin{minipage}[t]{72mm}
\resizebox{1.0\textwidth}{!}{\includegraphics{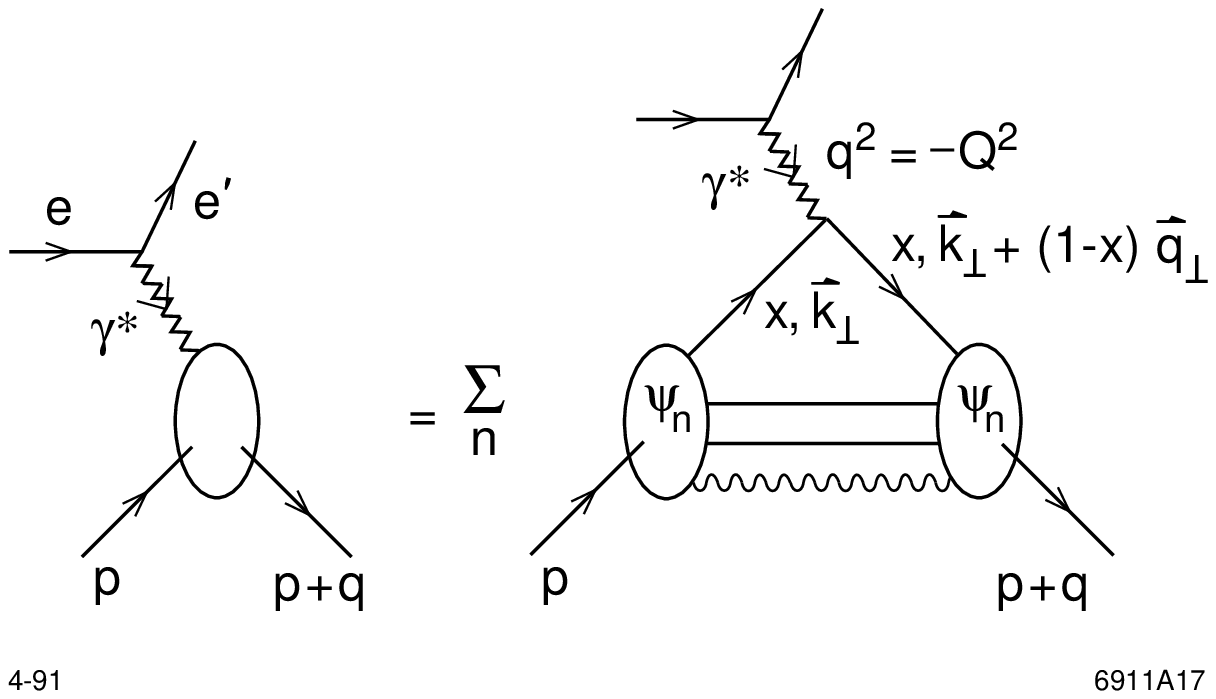}}
\caption{\label{fig:LLF}
   Calculation of the form factor of a bound state from the convolution
   of light-cone Fock amplitudes. The result is exact if one sums over
   all $\psi_n$.
}\end{minipage}\ \hfill
\begin{minipage}[t]{84mm}
  \resizebox{1.0\textwidth}{!}{\includegraphics{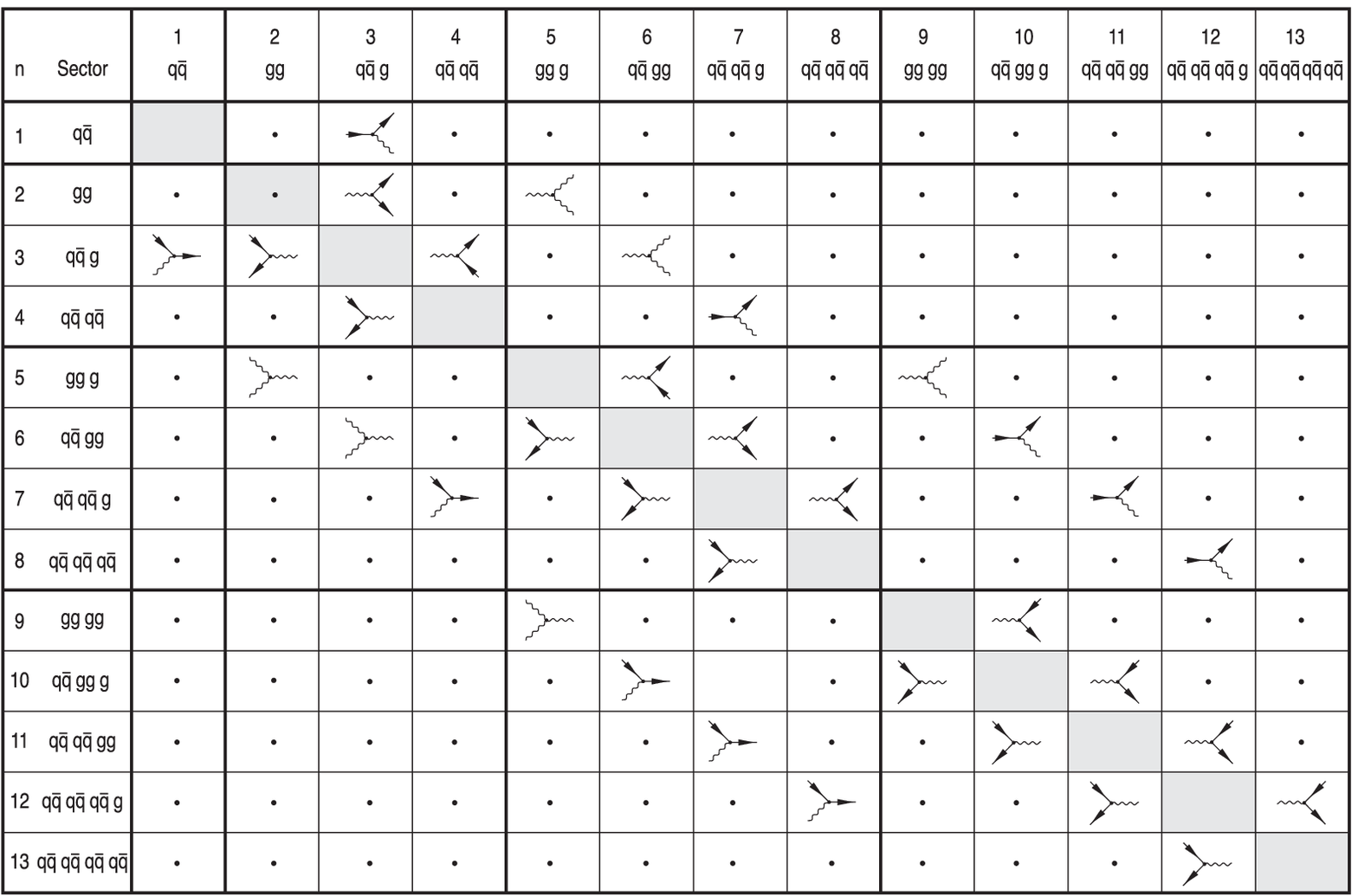}}
  \caption{\label{fig:2}  
     The Hamiltonian matrix for a meson. 
     The matrix elements are represented by energy diagrams. 
     Only vertex diagrams $V$ are shown.
     Zero matrices are marked by a dot ($\cdot$).
}\end{minipage}
\end{figure}
%-------------------------------------------------------------------

Let me sketch the method briefly. 
Details may be found in \cite{Pau98,Pau99b}.
DLCQ with its periodic boundary conditions has the advantage
that the LC-Hamiltonian is a matrix with a finite number of
Fock-space sectors, which we denumerate by $n$, 
with $1 < n\leq N$.  
The so called harmonic resolution $K= L P^+/(2\pi)$
acts as a natural cut-off of the particle number.
As shown in Figure~\ref{fig:2}, $K=3$ allows for $N=8$, 
and $K=4$ for $N=13$ Fock-space sectors, for example.
The Hamiltonian matrix is sparse: Most of the
matrix elements are zero,
particularly if one includes only the vertex interaction $V$.
For $n$ sectors, the eigenvalue problem in terms of block matrices reads 
\begin{eqnarray} 
   \sum _{j=1} ^{n} \langle i \vert H _n (\omega)\vert j \rangle 
                    \langle j \vert\Psi  (\omega)\rangle 
   =  E (\omega)\ \langle i \vert\Psi (\omega)\rangle 
,\qquad{\rm for}\ i=1,2,\dots,n
.\label{eq:2}\end{eqnarray} 
I can always invert the quadratic block matrix of the Hamitonian in 
the last sector to define the $n$-space resolvent $G _ n$, that is
\begin{eqnarray} 
    G _ n (\omega) =   {1\over \omega- H_n (\omega)} 
.\end{eqnarray}
Using $G _ n$, I can express the projection of the eigenfunction 
in the last sector by 
\begin{eqnarray} 
   \langle n \vert \Psi (\omega)\rangle   
   = G _ n (\omega) 
   \sum _{j=1} ^{n-1} \langle n \vert H _n (\omega)\vert j \rangle 
   \ \langle j \vert \Psi (\omega) \rangle 
,\end{eqnarray}
and substitute it in Eq.(\ref{eq:2}).
I then get an effective Hamiltonian where the number is sectors
is diminuished by 1:
\begin {equation}  
       H _{n -1} (\omega) =  H _n (\omega)
  +  H _n(\omega) G _ n  (\omega) H _n (\omega)
.\end {equation}
This is a recursion relation, which can repeated until one arrives
at the $q\bar q$-space.
The fixed point equation  $ E  (\omega ) = \omega $ determines then
all eigenvalues.

For the block matrix structure as in Figure~\ref{fig:2}, with its many 
zero matrices, the reduction is particularly easy and transparent.
For $K=3$ one has the following sequence of effective interactions:
\begin{eqnarray}
    H_8 = T_{8} 
,\quad
    H_7 = T_{7} + V G _8 V 
,\quad
     H_6 = T_{6} + V G _7 V 
,\quad
    H_5 = T_{5} + V G _6 V  
.\label{eq:6}\end{eqnarray}
The remaining ones get more complicated, \textit{i.e.}
\begin{eqnarray}
    H_4 
&=& T_{4} + V G _7 V + V G _7 V  G _6 V G _7 V 
,\\  
     H_3 
&=& T_{3} + V G _6 V + V G _6 V  G _5 V G _6 V + V G _4 V 
,\label{aeq:620}\\  
     H_2 
&=& T_{2} + V G _3 V + V G _5 V 
,\label{aeq:643}\\  
     H_1 
&=& T_{1} +  V G _3 V + V G _3 V  G _2 V G _3 V 
.\label{eq:9}\end{eqnarray}
For $K=4$, the effective interactions in Eq.(\ref{eq:6}) are different, 
see for example \cite{Pau99b}, but it is quite remarkable, that 
they are the same for the remainder, particularly Eq.(\ref{eq:9}).
In fact, the effective interactions in sectors 1-4 are independent of $K$:
The \emph{continuum limit $K\rightarrow\infty$ is then trivial},
and will be taken in the sequel. 

In the continuum limit, the effective interaction in the
$q\bar q$-space has thus two contributions:
A flavor-conserving piece $U_\mathit{eff-conser} = V G _3 V$
and a flavor-changing piece 
$U_\mathit{eff-change} = V G _3 V  G _2 V G _3 V$.
The latter cannot get active in flavor-off-diagonal mesons.
Notice that these expressions are an exact result.

\section{The eigenvalue equation in the $q\bar q$-space}
\label{sec:1}

After some approximations \cite{Pau98}, the effective one-body equation 
for flavor off-diagonal mesons
(mesons with a different flavor for quark and anti-quark),
becomes quite simple:
\begin{eqnarray} 
    &&M^2\langle x,\vec k_{\!\perp}; \lambda_{1},
    \lambda_{2}  \vert \psi\rangle  
    = \left[ 
    \frac{\overline m^2_{1} + \vec k_{\!\perp}^{\,2}}{x} +
    \frac{\overline m^2_{2} + \vec k_{\!\perp}^{\,2}}{1-x}  
    \right]\langle x,\vec k_{\!\perp}; \lambda_{1},
    \lambda_{2}  \vert \psi\rangle   
\label{eq:1}\\ 
    && - \frac{1}{3\pi^2}
    \sum _{ \lambda_q^\prime,\lambda_{2}^\prime} \!\int 
    \frac{ dx^\prime d^2 \vec k_{\!\perp}^\prime
    \,R(x',\vec k'_{\!\perp};\Lambda) }
    {\sqrt{ x(1-x) x^\prime(1-x^\prime)}}
    \frac{\overline\alpha}{\,Q^2} 
    \langle
    \lambda_{1},\lambda_{2}\vert S\vert 
    \lambda_{1}^\prime,\lambda_{2}^\prime\rangle
    \,\langle x^\prime,\vec k_{\!\perp}^\prime; 
    \lambda_{1}^\prime,\lambda_{2}^\prime  
    \vert \psi\rangle.  
\nonumber\end{eqnarray} 
Here, $M ^2$ is the eigenvalue of the invariant-mass squared. 
The associated eigenfunction $\psi\equiv\Psi_{q\bar q}$ 
is the probability amplitude 
$\langle x,\vec k_{\!\perp};\lambda_{1},\lambda_{2}\vert\psi\rangle$ 
for finding a quark with momentum fraction $x$, 
transversal momentum $\vec k_{\!\perp}$ 
and helicity $\lambda_{1}$,
and correspondingly the anti-quark with
$1-x$, $-\vec k_{\!\perp}$ and $\lambda_{2}$.
The $\overline m _1$  and $\overline m _2$ 
are (effective) quark masses  and 
$\overline\alpha$ is the (effective) coupling constant. 
The mean Feynman-momentum transfer of the quarks is
denoted by 
\begin{equation}
   Q^2 \equiv Q ^2 (x,\vec k_{\!\perp};x',\vec k_{\!\perp}') 
   = -\frac{1}{2}\left[(k_{1}-k_{1}')^2 + (k_{2}-k_{2}')^2\right]
,\end{equation}
and the spinor factor $S=S(x,\vec k_{\!\perp};x',\vec k_{\!\perp}')$ by 
\begin{equation}
   \langle\lambda_{1},\lambda_{2}\vert S\vert
   \lambda_{1}^\prime,\lambda_{2}^\prime\rangle =
   \left[ \overline u (k_{1},\lambda_{1})\gamma^\mu
   u(k_{1}^\prime,\lambda_{1}^\prime)\right] \, 
   \left[ \overline v(k_{2}^\prime,\lambda_{2}^\prime) \gamma_\mu 
    v(k_{2},\lambda_{2})\right] 
.\label{eq:3}\end{equation}
The regulator function $R(x',\vec k'_{\!\perp};\Lambda)$ 
restricts the range of integration as function of some mass scale 
$\Lambda$.
I happen to choose here a soft cut-off (see below),
in contrast to the previous sharp cut-off \cite{TriPau00}.  
Note that Eq.(\ref{eq:1}) is a fully relativistic equation.
I have derived the same effective interaction also with 
the method of Hamiltonian flow equations, see \cite{Pau00}.
 
The effective quark masses $\overline m _1$ and
$\overline m _2$ and  
the effective coupling constant $\overline\alpha$ 
depend, in general, on $\Lambda$.
In the spirit of renormalization theory they are renormalization
constants, subject to be determined by experiment,
and hence-forward will be denoted by $m _1$,
$m _2$, and $\alpha$, respectively.
In next-to-lowest order of approximation the coupling 
constant becomes a function of the momentum transfer, 
$\overline \alpha\longrightarrow\overline \alpha(Q;\Lambda)$, 
with the explicit expression given in \cite{Pau98}.

\section{The $\uparrow\downarrow$-model and its renormalization}
\label{sec:2}

It might be to early for solving Eq.(\ref{eq:1}) 
numerically in full glory like in Ref.\cite{TriPau00}.
Rather should I try to dismantle the equation of all irrelevant
details, and develop a simple model.

The quarks are at relative rest, 
when $\vec k _{\!\perp}= 0$ and 
$ x = \overline x \equiv m_1/(m_1+m_2)$. 
For very small deviations from these equilibrium values 
the spinor matrix is proportional to the unit matrix, with 
\begin{equation}
   \langle\lambda_1,\lambda_2\vert S\vert\lambda_1'\lambda_2'\rangle
   \sim 4 m_1 m_2 
   \ \delta_{\lambda_1,\lambda_1'}
  \ \delta_{\lambda_2,\lambda_2'}
,\end{equation}
for details see \cite{Pau00}.   
For very large deviations, particularly for
$\vec k_{\!\perp}^{\prime\,2} \gg \vec k_{\!\perp} ^{\,2}$,
holds  
\begin{equation}
   Q ^2 \simeq\vec k_{\!\perp}^{\prime\,2} 
,\hskip2em\mbox{ and } \hskip2em
   \langle\uparrow\downarrow\vert S\vert\uparrow\downarrow\rangle 
   \simeq 2\vec k_{\!\perp}^{\prime\,2}
.\end{equation}
Both extremes are combined in the ``$\uparrow\downarrow$-model''
\cite{Pau00}:
\begin{equation}
   \frac{S} {Q ^2} \equiv 
   \frac{4 m_1 m_2}{Q ^2} + 2 
   \Longrightarrow
   \frac{4 m_1 m_2}{Q ^2} + 2 R(\Lambda,Q) 
,\quad\textrm{with}\quad 
   R(\Lambda,Q) = \frac{\Lambda^2}{\Lambda^2+Q^2}
.\label{eq:4}\end{equation}
It interpolates between two extremes:
For small momentum transfer, the `2' generated by the hyperfine interaction
is unimportant and the dominant Coulomb aspects of the first term prevail.
For large momentum transfers the Coulomb aspects are
unimportant and the hyperfine interaction dominates.

%-------------------------------------------------------------------
\begin{figure} [t]
\begin{minipage}[t]{76mm}
  \resizebox{1.0\textwidth}{!}{\includegraphics{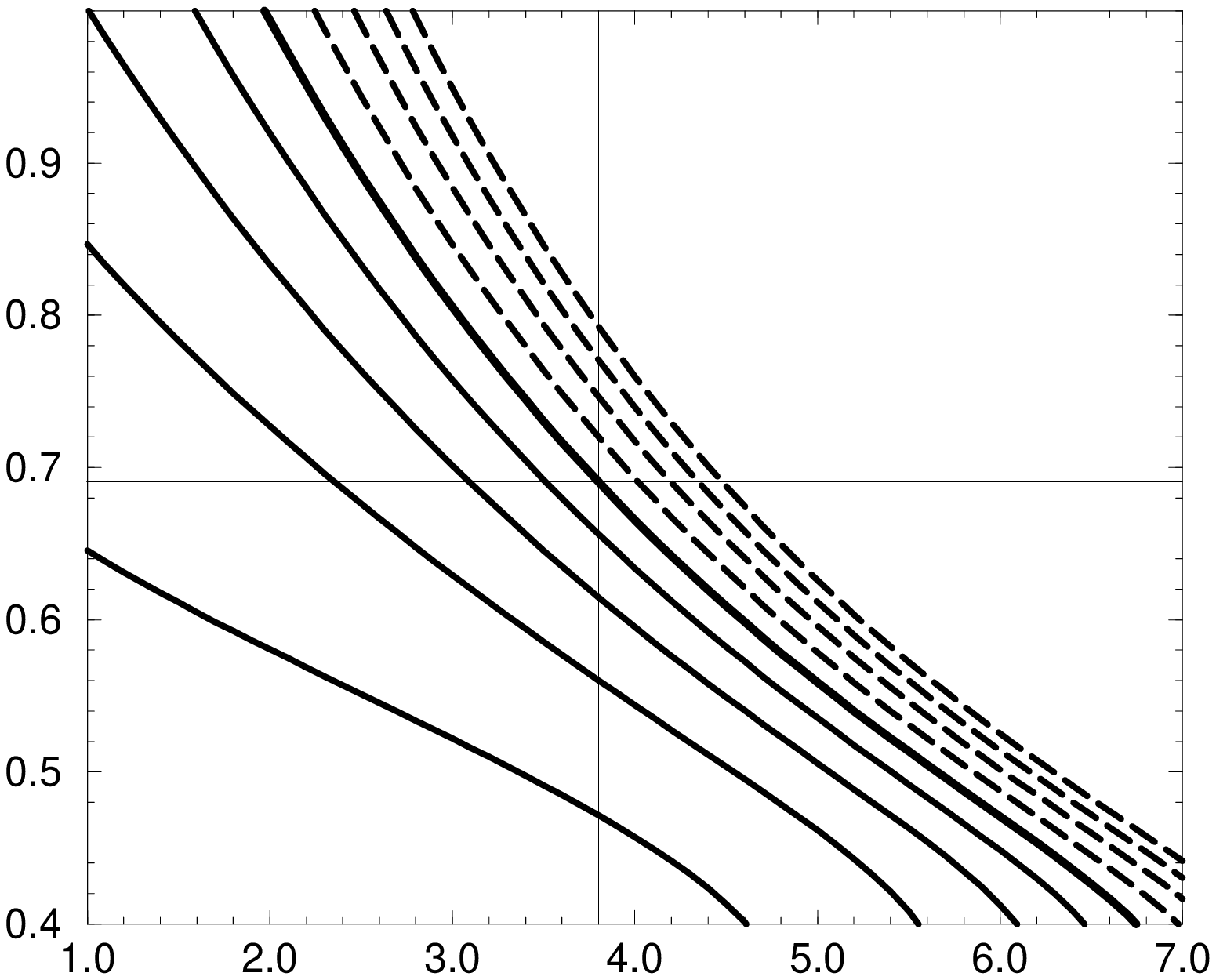}}
  \caption{\label{fig:3} 
   Nine contours $\alpha_n(\Lambda)$ are 
   plotted versus $\Lambda/\Delta$, 
   from bottom to top with $n=4,3,\cdots ,-3,-4$.
   The contours are obtained by 
   $M_0^2(\alpha,\Lambda) = n\Delta^2 + M_{\pi}^2$. 
   The thick contour $n=0$ refers to the pion with 
   $M_0^2=M_{\pi}^2$.
   Mass unit is $\Delta=350$~MeV.  
}
\end{minipage}\ \hfill
\begin{minipage}[t]{76mm}
  \resizebox{1.00\textwidth}{!}{\includegraphics{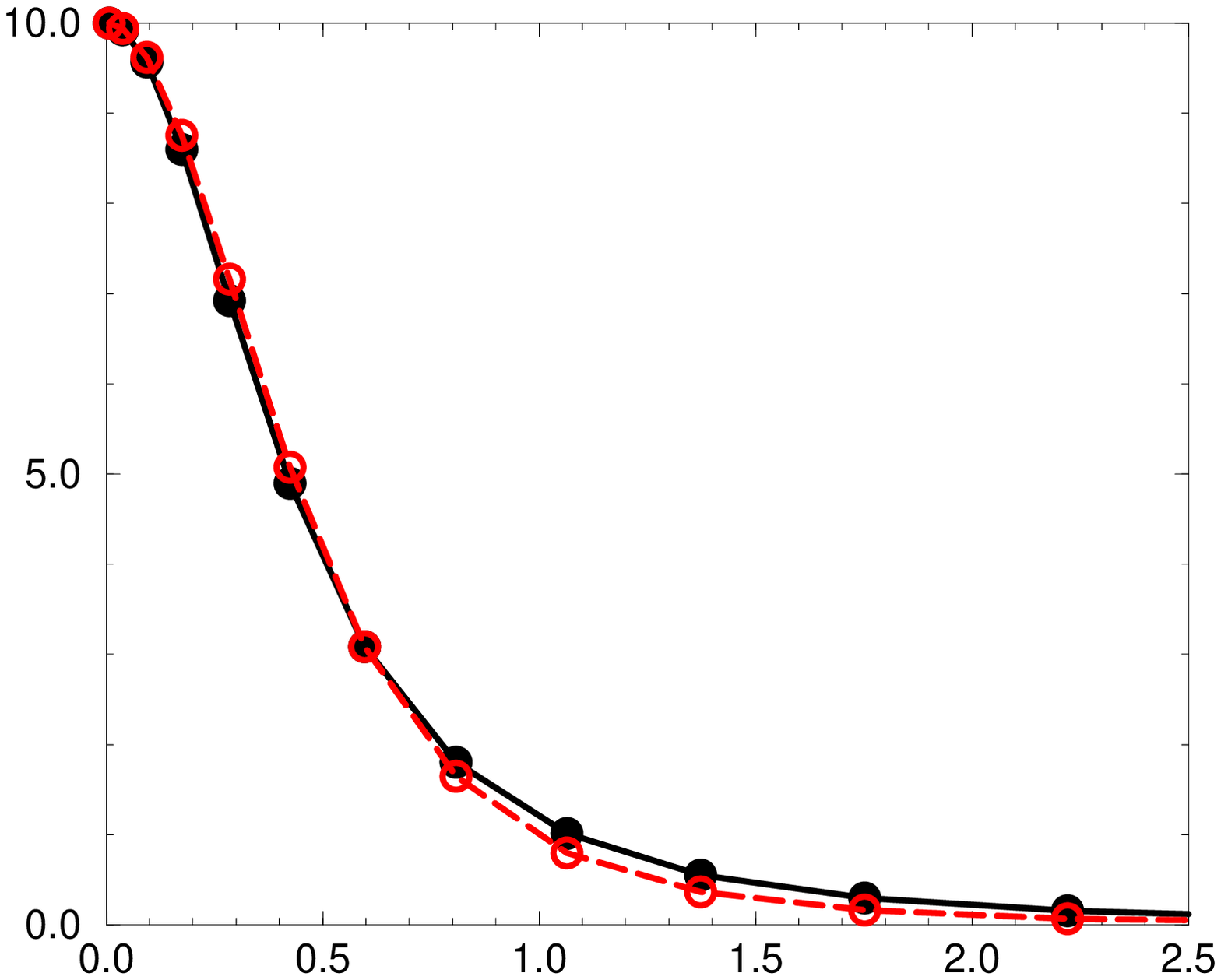}}
  \caption{\label{fig:6.1}   
   The pion wave function $\Phi(p)$ 
   is plotted versus $p/(1.338 m)$
   in an arbitrary normalization. 
   The filled circles indicate the numerical results,
   the open circles the analytical function $\Phi_a(p)$.
}\end{minipage}
\end{figure}
%-------------------------------------------------------------------

The model over-emphasizes many aspects: 
It neglects the momentum dependence of the Dirac spinors
and thus the spin-orbit interaction; it also neglects the
momentum dependence of the spin-spin interaction.
But the 2 creates havoc: Its Fourier transform is a Dirac-delta function 
with all its consequences in a bound-state equation.

Here is an interesting point: One is familiar
with field theoretic divergences like the effective masses 
and the effective coupling constant.
One is used less to ``divergences'' residing in a 
finite number 2.
They must be regulated also, and renormalized.

In consequence I replace Eq.(\ref{eq:1}) by
\begin{eqnarray} 
    && M^2
    \psi(x,\vec k_{\!\perp}) = \left[ 
    \frac{ m^2_{1} + \vec k_{\!\perp}^{\,2}}{x} +
    \frac{ m^2_{2} + \vec k_{\!\perp}^{\,2}}{1-x}  
    \right]\psi(x,\vec k_{\!\perp}) 
\nonumber\\&&-  
    \frac{\alpha}{3\pi^2} \!\int 
    \frac{ dx' d^2 \vec k_{\!\perp}'}
    {\sqrt{ x(1-x) x'(1-x')}}
    \left(\frac{4 m_1 m_2}{Q ^2} + 
    \frac{2\Lambda^2}{\Lambda^2+Q^2}\right)
    \psi(x',\vec k_{\!\perp}')
,\label{eq:20m}\end{eqnarray} 
where 
$\psi(x,\vec k_{\!\perp})\equiv
 \langle x,\vec k_{\!\perp}; \uparrow,\downarrow
 \vert \psi\rangle $.

For equal quark masses $m_1=m_2=m$, the eigenvalues depend now
on three parameters, the canonical $\alpha$ and $m$,
and the regularization scale $\Lambda$.
The dependence can be quite strong as seen in Figure~\ref{fig:3}.
There, the lowest mass-squared eigenvalue is plotted versus
$\alpha$ and $\Lambda$ for the fixed quark mass $m=406$~MeV.
 
Since $\Lambda$ is an unphysical parameter, 
its impact must be removed by renormalization.
Recently, much progress was made on this question \cite{FPF01,FFP01}:
Adding to $R(\Lambda,Q)$ a counterterm $C(\Lambda,Q)$ 
and requiring that the sum 
$\widetilde R(\Lambda,Q)=R(\Lambda,Q)+C(\Lambda,Q)$,
and thus $M^2(\Lambda;\alpha,m)$, be independent of $\Lambda$, 
determines $C(\Lambda,Q)$. One remains with 
$\widetilde R(\Lambda,Q)=\mu^2/(\mu^2+Q^2)$.
In line with renormalization theory, one then can go to the
limit $\Lambda\longrightarrow\infty$.
In turn, $\mu$ becomes one of the parameters of the theory 
to be determined by experiment.

\section{Determining the canonical parameters}

The theory has seven canonical parameters which have to be determined 
by experiment: $\alpha$, $\mu$ and the 5 flavor masses $m_f$
(if we disregard the top).
How can we determine them?

The problem is not completely trivial.
Let me restrict first to the light flavors. 
With $m_u=m_d=m$, one has 3 parameters, and in consequence needs
3 experimental data.
The pion mass $M_\pi=140$~MeV and the rho mass $M_\rho=768$~MeV
do not suffice. One needs a third datum, 
the mass of an exited pion, for example.

Since the mass of the excited pion $\pi^\pm$
is not known with sufficient experimental precision,
and since the $\uparrow\downarrow$-model might be to crude a model
to begin with, I choose here $m_u=m_d=406$~MeV and 
$M_{\pi^*}=M_\rho=768$~MeV, for no good reason other than convenience.
These assumptions are less stringent than they sound,
by two reasons. First, the rho has a mass less than $2m$ 
and should be a true bound state.
Second, the Yukawa potential in Eq.(\ref{eq:20m}) 
acts like a Dirac-delta function in pairing theory for example:
it pulls down essentially one state, the pion,
but leaves the other states unchanged.

We thus remain with the two parameters $\alpha$ and $\mu$.
Each of the two equations, 
$M_0^2(\alpha,\mu)=M_\pi^2$ and 
$M_1^2(\alpha,\mu)=M_{\pi^*}^2$ 
determine a function $\alpha(\mu)$.
Their intersection point determines the required solution,
which is $\alpha=0.761$ and $\mu=1.15$~GeV \cite{FPF01}.
These differ marginally from the previous analysis \cite{Pau00},
with $\mu=1.33$~GeV, for which Figure~\ref{fig:3} yields $\alpha=0.6904$.
Once I have the up and down mass, 
the strange, charm and bottom quark mass can be determined
by reproducing the masses of the 
$K,^-$ $D^0$ and $B,^-$ respectively. 
The parameters in the $\uparrow\downarrow$-model 
can thus be taken as 

\begin{tabular}{ccccccc} 
 $\alpha$ & $\mu$   & $m_u=m_d$& $m_s$  & $m_c$   & $m_b$   \\
  0.6904  & 1.33 GeV& 406 MeV  & 508 MeV& 1666 MeV& 5054 MeV.    
\end{tabular}
\ \hfill(I)

%-------------------------------------------------------------------
\begin{table} [t]
\begin{minipage}[t]{75mm}
\begin{tabular}{c|rrrrrr} 
%   \rule[-1em]{0mm}{1em}
     & $\overline u$ & $\overline d$ 
     & $\overline s$ & $\overline c$ & $\overline b$ \\ \hline
%   \rule[1em]{0mm}{0.5em}
    $u$ &         &      768&      871&     2030&     5418 \\
    $d$ &      140&         &      871&     2030&     5418 \\
    $s$ &      494&      494&         &     2124&     5510 \\
    $c$ &     1865&     1865&     1929&         &     6580 \\
    $b$ &     5279&     5279&     5338&     6114&          
\end{tabular}
\caption{\label{tab:2.2}  
   The calculated mass eigenvalues in MeV. 
   Those for singlet-1s states are given in the lower,
   those for singlet-2s states in the upper triangle.}
\end{minipage}\ \hfill
\begin{minipage}[t]{75mm}
\begin{tabular}{c|rrrrrr} 
     & $\overline u$ & $\overline d$ 
     & $\overline s$ & $\overline c$ & $\overline b$ \\ \hline
 $u$ &      & 768  & 892  & 2007 & 5325 \\ 
 $d$ & 140  &      & 896  & 2010 & 5325 \\ 
 $s$ & 494  & 498  &      & 2110 &  --- \\ 
 $c$ & 1865 & 1869 & 1969 &      &  --- \\ 
 $b$ & 5278 & 5279 & 5375 &  --- &      \\ 
\end{tabular}
\caption{\label{tab:1.2}  
   Empirical masses of the flavor-off-diagonal physical mesons in MeV.
   Vector mesons are given in the upper, scalar mesons
   in the lower triangle.}
\end{minipage}
\end{table}
%-------------------------------------------------------------------

\section{The masses of the physical mesons}

Solving Eq.(\ref{eq:20m}) with the parameters of Eq.(I) generates the 
mass$^2$-eigenvalues of all flavor off-diagonal pseudo-scalar mesons. 
They are compiled in Table~\ref{tab:2.2}. 
The corresponding wave functions are also available, but not shown here.
In view of the simplicity of the model, the agreement with the empirical 
values \cite{PDG98} in Table~\ref{tab:1.2} is remarkable. 
The mass of the first excited states in Table~\ref{tab:2.2} 
correlates astoundingly well with the experimental mass of the 
pseudo-vector mesons, as given in Table~\ref{tab:1.2}.
Notice that all numbers in Tables~\ref{tab:2.2}
and \ref{tab:1.2} are rounded for convenience. 

Since the $\uparrow\downarrow$-model in Eq.(\ref{eq:20m}) 
does not expose confinement one should emphasize that the difference 
between the physical meson masses in Table~\ref{tab:2.2}
amd the sum of the bare quark masses is larger than a pion mass.
One could call this a kind of practical confinement.

%-------------------------------------------------------------------
\begin{table} [t]
\begin{minipage}[t]{78mm}
\begin{center}
\begin{tabular}{c|rrr} 
  \rule[-1em]{0mm}{1em}
  { }      & $M_{f\bar f}$ &    $M$ & $M_\mathrm{exp}$ \\ \hline
  \rule[1em]{0mm}{0.5em}
  $\pi^0$  &   140         &    140 &  135 \\ 
  $\eta $  &   140         &    485 &  549 \\ 
  $\eta'$  &   661         &    958 &  958 \\ 
  $\eta_c$ &  2870         &   2915 & 2980 \\ 
  $\eta_b$ &  8922         &   8935 &  --- \\ 
\end{tabular}
\end{center}
\caption{\label{tab:3} 
   Flavor-diagonal mass eigenvalues in the FM-model for pseudo-scalar 
   mesons with the parameter $a = (491\mbox{ MeV})^2$.}
\end{minipage}\ \hfill
\begin{minipage}[t]{78mm}
\begin{center}
\begin{tabular}{c|rrr} 
  \rule[-1em]{0mm}{1em}
  { }      & $M_{f\bar f}$ &    $M$ & $M_\mathrm{exp}$ \\ \hline
  \rule[1em]{0mm}{0.5em}
  $\rho^0$   &  768 &     768 &  768 \\ 
  $\omega$   &  768 &     832 &  782 \\ 
  $\Phi  $   &  973 &    1019 & 1019 \\ 
  $J/\Psi $  & 3231 &    3242 & 3097 \\ 
  $\Upsilon$ & 9822 &    9825 & 9460 \\ 
\end{tabular} 
\end{center}
\caption{\label{tab:4} 
   Flavor-diagonal mass eigenvalues in the FM-model for pseudo-vector 
   mesons with the parameter $a = (255\mbox{ MeV})^2$.}
\end{minipage} 
\end{table}
%-------------------------------------------------------------------
What about the flavor diagonal mesons?-- They cannot be a solution
to Eqs.(\ref{eq:1}) or (\ref{eq:20m}), since the flavor-changing 
piece of the full effective interaction can generate matrix elements 
between different flavors. Thus far the precise structure of the 
flavor changing part 
%A flavor-conserving piece $U_\mathit{eff-conser} = V G _3 V$
$U_\mathit{eff-change} = V G _3 V  G _2 V G _3 V$ 
has not been analyzed in detail, 
because it requires a considerable effort.

Rather, the following flavor-mixing model (FM-model \cite{Pau01c}) 
has been investigated.
In the FM-model, the full effective Hamiltonian including its flavor
mixing is reduced to the lowest $f\bar f$-states, \textit{i.e.} to
\begin{eqnarray}
   \langle f\bar f \vert H_\mathit{eff}\vert f'\bar f '\rangle =
   \langle \psi_{f\bar f} \vert T + V G _3 V + V G _3 V  G _2 V G _3 V
   \vert \psi_{f'\bar f'} \rangle = 
   M_{f f'} ^2\ \delta _{f f'} + a
.\end{eqnarray}
Conceptually, it is important that $M_{f f'} ^2$ is the eigenvalue of 
Eq.(\ref{eq:20m}). The flavor-mixing matrix element
$ a = \langle \psi_{f\bar f} \vert V G _3 V  G _2 V G _3 V
      \vert \psi_{f'\bar f'} \rangle $ 
depends on the flavors and could be calculated
with a solution of Eqs.(\ref{eq:1}) or (\ref{eq:20m}).
In the crude FM-model, however, it is treated as a flavor-independent 
parameter to be fixed by experiment.
For 5 flavors one faces thus the numerical diagonalization 
of a $5\times 5$ matrix.

The parameter $a$ for pseudo-scalar mesons was fitted to the mass
of the $\eta'$, and for pseudo-vector mesons to the $\Phi$,
with the results compiled in Tables~\ref{tab:3} and  \ref{tab:4} . 
Three facts, however, one gets for free:
First, the $\pi^0$ is degenerate in mass with $\pi^\pm$,
as well as the $\rho^0$ with $\rho^\pm$. 
That they form isospin-triplets is a non-trivial aspect of QCD.
Second, both the $\eta$--$\eta'$ and the $\omega$--$\Phi$ splitting
are in the right bull park. 
Third, that the wave functions of the $\pi^0,\eta$ and $\eta'$ 
have very much SU(3)-character \cite{Pau01c} is even less trivial 
from the point of view of QCD.

\section{The wavefunction of the pion}
For carrying out this programme in practice, 
I need an efficient tool for solving Eq.(\ref{eq:20m}). 
Such one has been developed recently \cite{Pau00}. 
I outline in short the procedure for the special case
$m_1=m_2=m$.
I change integration variables from $x$ to $k_z$ 
by substituting
\begin{equation}
   x(k_z) = \frac{1}{2} +
   \frac{k_z}{2\sqrt{m^2 + \vec k_{\!\perp}^{\,2} + k_z^2}}
,\hskip3em
   \psi(x,\vec k _{\!\perp}) =
   \frac{\sqrt{1+(\vec k_{\!\perp}^{\,2} + k_z^2)/m^2}}{\sqrt{x(1-x)}}
   \ \phi(k_z,\vec k _{\!\perp})
.\label{peq:6}\end{equation}
The variables $(k_z,\vec k_{\!\perp})$ are 
collected in a 3-vector $\vec p$ and Eq.(\ref{eq:20m}) becomes 
\begin{eqnarray}
   \left[M^2-4m^2-4\vec p^{\,2}\right]\phi(\vec p) &=& 
   -\frac{4\alpha}{3\pi^2}\int\!\! d^3\vec p\,' 
   \left(\frac{2m}{(\vec p - \vec p\,')^{\,2}} + \frac{1}{m}
   \frac{\mu^2}{\mu^2+(\vec p - \vec p\,')^{2}} \right)
   \phi(\vec p\,')
.\end{eqnarray}
For the present purpose it suffices to restrict to 
spherically symmetric $s$-states $\phi(\vec p)=\phi(p)$
and to apply Gaussian quadratures with 16 points.
On an alpha work station it takes a couple of micro-seconds 
to solve this equation for a particular case.
The resulting numerical wavefunction $\phi(p)$ is displayed in 
Figure~\ref{fig:6.1} and compared with 
\begin{equation}
  \Phi_a(p) = \mathcal{N}
  \left(1+\displaystyle\frac{p^2}{p_a^2}\right)^{-2} 
  ,\hskip3em\mathrm{with}\quad
  p_a=1.338\ m
.\label{peq:12}\end{equation}
Such an analytical form is convenient in many applications.
For example, the light-cone wavefunction $\psi(x,\vec k_{\!\perp})$ 
can be obtained in closed form by Eq.(\ref{peq:6}), \textit{i.e.}
\begin{eqnarray}
   \psi(x,\vec k _{\!\perp}) =
   \frac{\mathcal{N}}{\sqrt{x(1-x)}} \frac
   {\left(1+\displaystyle
   \frac{m^2\left(2x-1\right)^2+\vec k_{\!\perp}^{\,2}}
   {4x(1-x)\ m^2} \right)^{\frac{1}{2}}}
   {\left(1+\displaystyle
   \frac{m^2\left(2x-1\right)^2+\vec k_{\!\perp}^{\,2}}
   {4x(1-x)\ p_a^2} \right)^2} 
.\label{peq:15}\end{eqnarray}
I can use this to calculate the form factor from Eq.(\ref{eq:b1}), 
and thus the exact root-mean-square radius \cite{PaM01},
even in closed form with 
$\langle r^2\rangle = -6\left.{dF_2(Q^2)}/{dQ^2}\right\vert_{Q^2=0}$:
\begin{eqnarray}
   \langle r^2\rangle  &=& 
   \frac{3}{4p_a^2}\ \frac
   {34 +37s^2 -41s^4 +15s^6 + 3b(s)(- 8 -16s^2 +21s^4 -17s^6+5s^8)}
   {5(s^2-1)[4-4s^2+3s^4+3b(s)s^4(-2+s^2)]}
,\end{eqnarray}
with $s={m}/{p_a}$ and the abbreviation 
$b(s)= \mathrm{arctan}( \sqrt{s^2-1})/\sqrt{s^2-1}$.
The size of the $q\bar q$  wavefunction is thus
$\langle r^2\rangle ^{\frac{1}{2}} = 0.33$~fm, 
half as large as the empirical value   
$\langle r^2\rangle_{\mathrm{exp}} ^{\frac{1}{2}}=0.67$~fm.

This completes the goal: I have a pion with the correct mass, 
and I have an analytic expression for its light-cone wave function.
Eq.(\ref{peq:15}) could be used thus as a baseline for calculating 
the higher Fock-space amplitudes, as explained in \cite{Pau99b}.

It might well be that the wavefunction in Eq.(\ref{peq:15}) 
is consistent with Ashery's experiment \cite{Ash00}.

\section{Conclusions}

The proposed pion of the $\uparrow\downarrow$-model is rather
different from the pions in the literature.
I have found no evidence that the vacuum condensates are important,
but I conclude that the pion is describable by a QCD-inspired theory:
The very large coupling constant in conjunction 
with a very strong hyperfine interaction
makes it a ultra strongly bounded system of constituent quarks.
More then 80 percent of the constituent quark mass is eaten up by 
binding effects.
No other physical system has such a property.

The effective Bohr momentum of the constituents in the pion
turns out as $p_a=1.338\,m$, see Eq.(\ref{peq:12}).
The mean momentum of the constituents is thus 40 percent 
larger than their mass, which means that 
they move highly relativistically quite in contrast with the 
constituents of atoms or nuclei.
No wonder that potential models thus far have failed for the pion.
One might mention that lattice gauge calculations
use all the computer power in this world to generate the potential energy
of the quarks and then one uses a \emph{non-relativistic}
Schr\"odinger equation to calculate the bound states.

All this is to be confronted with the present oversimplified 
$\uparrow\downarrow$-model,
which however has the virtue to calculate the pion and other physical mesons
by a covariant and relativistically correct theory.
To the best of my knowledge there is no other model which can describe
\emph{ all mesons} quantitatively from the $\pi$ up to the $\Upsilon$
from a common point of view, which here is QCD.

\end{document}